\documentclass[aps,prd,showpacs,onecolumn,preprintnumbers,notitlepage,superscriptaddress,
nofootinbib,amsfont,amssymb,10pt,singlespacing] {revtex4-1}
\usepackage{amsmath,bm}
\usepackage{times}
\usepackage{braket}
\usepackage{color,graphicx}
\usepackage{slashed}
\usepackage{hyperref}

\newcommand{\beq}{\begin{eqnarray}}
\newcommand{\eeq}{\end{eqnarray}}

\def\lsim{ {\ \lower-1.2pt\vbox{\hbox{\rlap{$<$}\lower6pt\vbox{\hbox{$\sim$}
}}}\ } }
\def\gsim{ {\ \lower-1.2pt\vbox{\hbox{\rlap{$>$}\lower6pt\vbox{\hbox{$\sim$}
}}}\ } }
\begin{document}

\title{Parton distribution of nucleon and nuclear EMC effect in a statistical model}
%%%==================================================================
\author{Xian-Qiao Yu}
\email[Electronic address:]{yuxq@swu.edu.cn}
\affiliation{School of Physical Science and Technology,\\ Southwest University, Chongqing 400715, China}
%%%%%%%%%%%%%%%%%%%%%%%%%%%%%%%%%%%%%%%%%%%%%%%%%%%%%%%%%%%%%%%%%%
\begin{abstract}
We study the parton distribution of nucleon and nuclear EMC effect in a statistical model. We find when we choose the parameters appropriately,
the predictions given by pure statistical laws can fit the experimental data well in most range of $x$, this reveal statistical law play an
important role in the parton distribution of nucleon.
\end{abstract}
%%%%%%%%%%%%%%%%%%%%%%%%%%%%%%%%%%%%%%%%%%%%%%%%%%%%%%%%%%%%%%

\maketitle

%
%%%
%%%%%%%%%%%%%%%%% I. INTRODUCTION %%%%%%%%%%%%%%%%%%%%%%%%%%%%%%%%
%%%
%

\section{Introduction}

Deep inelastic lepton scattering provide a good platform to study the internal structure of nucleon, such as what it is consists of and how its constituent
distribute in momentum space? In a simple quark model, nucleon is consist only of three quarks. This simple quark model has successfully explained a few of
 basic properties of nucleon---including its mass, spin and parity, etc. However, it is not enough for explaining the experimental data from deep inelastic lepton
 off nucleon. In order to solve this problem, Feynman put forward a very successful model--parton model of nucleon\cite{Feynman}, now we have known that the partons
 in nucleon including valence quarks, sea quarks and gluon, which is far more complicate than the simple quark model. The structure function, which contains the
 information of nucleon structure, can be easily presented by the parton's momentum distribution in a Infinite Momentum Frame(IMF)
 \begin{equation}
F_{2}(x)=\sum_{i}e_{i}^{2}xf_{i}(x), \label{strf}
\end{equation}
where $e_{i}$ is the charge of the ith parton and $f_{i}(x)$ the ith parton's momentum distribution function. Now, the key question is how to compute the
parton's momentum distribution function? Strict
analysis of the momentum distribution of nucleon constituents should start from
Quantum Chromodynamics(QCD). Because the complication of QCD non-perturbation, it is very difficult to compute the distribution
function from the first principle of QCD. Considering the partons in nucleon may obey some determined statistical law, I shall make a attempt to calculate
the distribution function from statistical physics in this paper, as I shall show in the following.

%%%%%%%%%%%%%%%%%% II. Formalism %%%%%%%%%%%%%%%%%%%%%%%%%%%%%%%%
%%%

\section{ The valence quark distribution in a free nucleon }\label{sec:vqd}

In the IMF, the interaction between the partons becomes ``slowly'', and hence the partons in nucleon can be safely
regarded as ideal gas and they should obey statistical laws. We suppose the valence quarks in nucleon obey Boltzmann distribution. Under equilibrium state,
 Boltzmann  distribution gives the probability of finding a valence quark in the range of $\xi\rightarrow\xi+d\xi$
\begin{equation}
f(\xi)d\xi=\frac{\beta}{1-e^{-\beta}}e^{-\beta\xi}d\xi, \label{ovqd}
\end{equation}
which is one valence quark's momentum distribution function and $\beta$ is determined by
\begin{equation}
\frac{1-e^{-\beta}(1+\beta)}{\beta(1-e^{-\beta})}=\frac{\xi_{v}}{3}, \label{betae}
\end{equation}
where$\xi_{v}$
is the ratio of total valence quarks's momentum to nucleon's momentum($\xi_{v}=\frac{P_{v}}{P_{N}}$).

\begin{figure}[!!htb]
\centering
\begin{tabular}{l}
\includegraphics[width=0.55\textwidth]{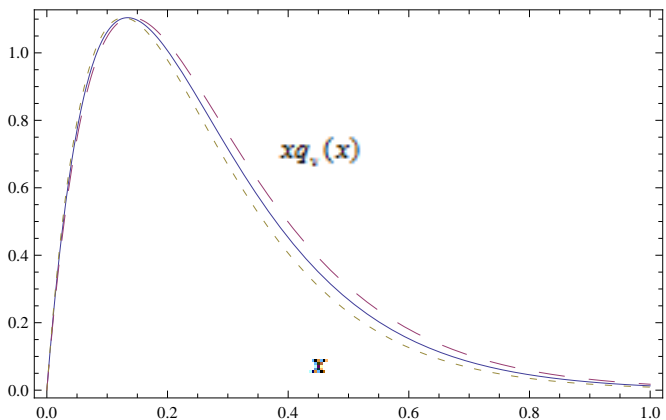}
\end{tabular}
\caption{The valence quarks momentum distribution function in a free nucleon, where solid line for $\xi_{v}$=0.4, dotted line for $\xi_{v}$=0.38
and dashed line for $\xi_{v}$=0.42.}
  \label{fig:fig1}
\end{figure}

Here, we have regarded up quark and down quark in nucleon as identical particle. Because proton consists two up quarks and one down quark, we have
$u_{v}(x)=2d_{v}(x)$ for proton, here $x$ is Bjorkon scaling($x=Q^{2}/2M_{N}\nu$). In this paper, we shall use $x$ and $\xi$ exchangeable, for $x=\xi$ in IMF.
 The deviation from the above relation is observed in experiment mainly because of the mass difference between up quark and down quark. From Eqs.(\ref{ovqd}) and (\ref{betae}),
 we can see that the valence quark momentum distribution in nucleon is only determined by a free parameter$\xi_{v}$. We draw the contribution from valence quarks $xq_{v}(x)$ in
 a free nucleon in Fig.~\ref{fig:fig1} for three different value of $\xi_{v}$.
%%%%=============================================================
%%%%==============================================================

Our calculations show that the valence quark's contribution is not sensitive to its momentum ratio $\xi_{v}$, but sensitive to the number of valence
quarks. It implies that all the baryons's momentum distribution function should be similar, but there should be significant differences between
baryons's momentum distribution function and mesons's momentum distribution function.

\section{The sea quark distribution in a free nucleon }\label{sec:sqd}

In nucleon, a gluon can annihilate into a pair of sea quarks($g\rightarrow q_{s}\bar{q}_{s}$); meanwhile, a pair of sea quarks also can be annihilate into
a gluon($q_{s}\bar{q}_{s}\rightarrow g$). In equilibrium state, gluon and sea quarks are in phase equilibrium, the two phase's chemical potential equal($\mu_{g}=\mu_{s}$).
The number of sea quark is undetermined, but only N=0, 2, 4, 6, 8, $\cdots \cdots$ are allowed. Suppose the ratio of total longitudinal momentum of sea quarks to
the total nucleon momentum is $\xi_{s}$($\xi_{s}=P_{s}/P_{N}$), grand canonical distribution gives the probability of there is N sea quarks

 \begin{eqnarray}
 \frac {e^{-\alpha N}\sum\limits_{s} e^{-\beta\xi_{s}}}{\sum\limits_{N}\sum\limits_{s}e^{-\alpha N-\beta\xi_{s}}},\label{nnsqp}
\end{eqnarray}
where $\alpha$ is a constant which is dependent on the chemical potential $\mu_{g}$ or $\mu_{s}$, the sum on $s$ means it goes over all the
possible microscopic states.

When N=2, from the discussions in Section(\ref{sec:vqd}), we get one sea quark momentum distribution

\begin{equation}
f(\xi)d\xi=\frac{\beta_{2}}{1-e^{-\beta_{2}}}e^{-\beta_{2}\xi}d\xi, \label{2osqd}
\end{equation}
and $\beta_{2}$ is determined by
\begin{equation}
\frac{1-e^{-\beta_{2}}(1+\beta_{2})}{\beta_{2}(1-e^{-\beta_{2}})}=\frac{\xi_{s}}{2}. \label{2sbetae}
\end{equation}

When N=4, one sea quark momentum distribution function is

\begin{equation}
f(\xi)d\xi=\frac{\beta_{4}}{1-e^{-\beta_{4}}}e^{-\beta_{4}\xi}d\xi, \label{4osqd}
\end{equation}
and $\beta_{4}$ is determined by
\begin{equation}
\frac{1-e^{-\beta_{4}}(1+\beta_{4})}{\beta_{4}(1-e^{-\beta_{4}})}=\frac{\xi_{s}}{4},\label{4sbetae}
\end{equation}
and so on$\cdots\cdots$

So, the total sea quarks momentum distribution function is
 \begin{eqnarray}
f_{s}(\xi)d\xi=\frac {e^{-2\alpha}\sum\limits_{s} e^{-\beta\xi_{s}}}{\sum\limits_{N}\sum\limits_{s}e^{-\alpha N-\beta\xi_{s}}}\frac{2\beta_{2}}{1-e^{-\beta_{2}}}e^{-\beta_{2}\xi}d\xi \nonumber \\
+\frac {e^{-4\alpha}\sum\limits_{s} e^{-\beta\xi_{s}}}{\sum\limits_{N}\sum\limits_{s}e^{-\alpha N-\beta\xi_{s}}}\frac{4\beta_{4}}{1-e^{-\beta_{4}}}e^{-\beta_{4}\xi}d\xi \nonumber \\
+\frac {e^{-6\alpha}\sum\limits_{s} e^{-\beta\xi_{s}}}{\sum\limits_{N}\sum\limits_{s}e^{-\alpha N-\beta\xi_{s}}}\frac{6\beta_{6}}{1-e^{-\beta_{6}}}e^{-\beta_{6}\xi}d\xi+\cdots\cdots\label{sqd}
\end{eqnarray}

There is a difficulty in calculating the probability of N sea quarks
\begin{eqnarray}
 \frac {e^{-\alpha N}\sum\limits_{s} e^{-\beta\xi_{s}}}{\sum\limits_{N}\sum\limits_{s}e^{-\alpha N-\beta\xi_{s}}}, \label{2nnsqp}
\end{eqnarray}
for the microscopical state number of macroscopical state $\xi_{s}$ varies with N in general. Supposing the difference
is small and we can compute formula(\ref{2nnsqp}) approximately
\begin{eqnarray}
 \frac {e^{-\alpha N}\sum\limits_{s} e^{-\beta\xi_{s}}}{\sum\limits_{N}\sum\limits_{s}e^{-\alpha N-\beta\xi_{s}}}
 \approx\frac{e^{-\alpha N}}{\sum\limits_{N}e^{-\alpha N}}=e^{-\alpha N}(1-e^{-2\alpha}). \label{3nnsqp}
\end{eqnarray}

\begin{figure}[!!htb]
\centering
\begin{tabular}{l}
\includegraphics[width=0.55\textwidth]{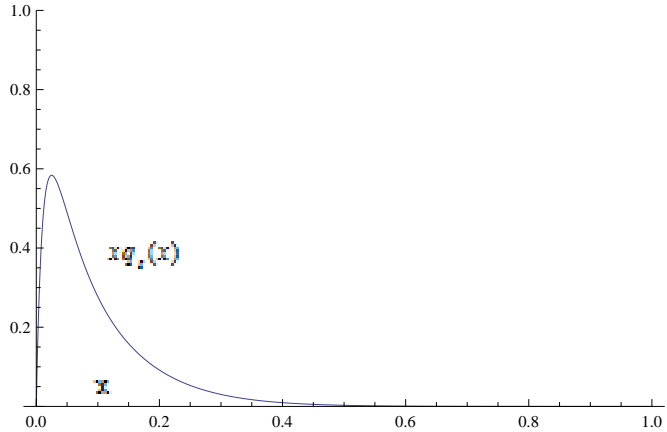}
\end{tabular}
\caption{The sea quarks momentum distribution function in a free nucleon}
  \label{fig:fig2}
\end{figure}

In this approximation, we draw the contribution from sea quarks $xq_{s}(x)$ in
a free nucleon in Fig.~\ref{fig:fig2},
%%%%=============================================================
%%%%==============================================================
where we have taken $\xi_{s}=0.14$, $\alpha=0.36$ and have calculated the contributions up to N=40, more sea quarks's contribution is very small and can be safely omitted.

We find that the sea quarks's contribution is not sensitive to the parameters $\xi_{s}$ and $\alpha$, the calculations indicate that its contributions
are mainly concentrated in the small $x$ region, as is shown in Fig.~\ref{fig:fig2}.

\section{The structure function of a free proton }\label{sec:freensf}

\begin{figure}[!!htb]
\centering
\begin{tabular}{l}
\includegraphics[width=0.55\textwidth]{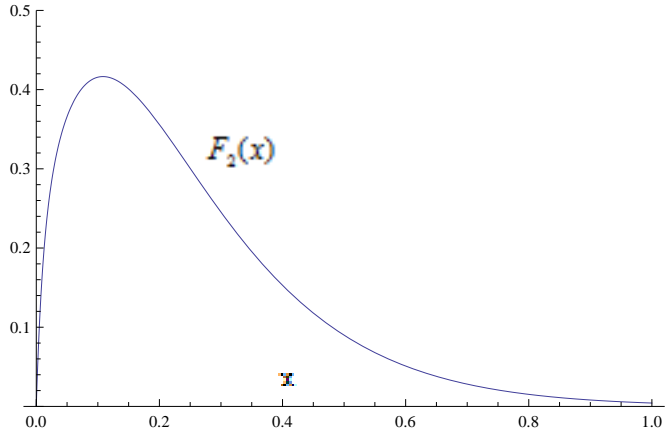}
\end{tabular}
\caption{The structure function $F_{2}(x)$ of a free proton}
  \label{fig:fig3}
\end{figure}

In this section, we derive the structure function $F_{2}(x)$ of
a free proton. From Eq.(\ref{strf}), we obtain the structure function $F_{2}(x)$ for a proton
\begin{equation}
F_{2}^{p}(x)=\frac{1}{3}xq_{v}(x)+\frac{2}{9}xq_{s}(x),\label{strfp}
\end{equation}
where $xq_{v}(x)$ and $xq_{s}(x)$ represent respectively the total valence quarks's contribution and total sea quarks's contribution, as are shown in Fig.~\ref{fig:fig1} and Fig.~\ref{fig:fig2}.
In this statistical model, there are only three free parameters $\xi_{v}$, $\xi_{s}$ and $\alpha$. Experiment
shows that the contribution of sea quarks is about ten's percent. Taking $\xi_{v}=0.40$, $\xi_{s}=0.14$ and $\alpha=0.36$,
we get the structure function $F_{2}(x)$ of a free proton in Fig.~\ref{fig:fig3}, where the contribution of sea quarks we have calculated up to N=40,
for the reason we have mentioned above. We find when $x=0.1$, $F_{2}^{p}(x)$ is about 0.4, which is in accordance with the experimental data. 
When $x$ continue to increase, the structure function $F_{2}^{p}(x)$ decreases, a little slower than the experimental data. In small $x$ range, 
our curve drops, the reason can be easily seen from the right side of Eq.(\ref{strf}). However, the
 experimental results for proton structure function $F_{2}(x)$ show a strong rise towards small $x$\cite{Adloff,Aid,Derrick}, which disagrees with the
 predictions of Fig.~\ref{fig:fig3}, it reveals the limitations of pure statistical law.

%%%%=============================================================
%%%%==============================================================

\section{The EMC effect} \label{sec:EMCE}
In the above paper, we have discussed the parton distribution of a free nucleon in the statistical model.
If we measure the parton distribution of a nucleon which is confined in a nuclei, defining
 \begin{eqnarray}
 F_{2}^{A}(x)=[ZF_{2}^{p}(x)+NF_{2}^{n}(x)]/A,\label{ansf}
\end{eqnarray}
where A is the atomic number with Z protons and N neutrons, denoting structure function of deuterium $F_{2}^{D}
=(F_{2}^{p}(x)+F_{2}^{n}(x))/2$. Defining the ratio R(x)

\begin{equation}
R(x)\equiv F_{2}^{A}(x)/F_{2}^{D}(x), \label{ratio}
\end{equation}
one would expect that $R(x)\approx1$. However, in 1983 the European Muon Collaboration(EMC) discovered that
R(x) deviates from one significantly for iron\cite{Aubert}, later confirmed by more extensive measurements at SLAC\cite{Gomez}. This
phenomenon is known as the `EMC effect'. Since then how to explain EMC effect has become a attracted subject
in theoretical physics and much effort had been made to understand the underlying physics of the effect\cite{Close,Nachtmann,Canal,Staszel,Qiu}. In this paper,
we give a explanation for EMC effect from statistical physics's view. We will find in most range it could fit the experimental
data well, this implies that statistical principle play an important role in the parton distribution of nucleon(not only
the free nucleon, but the confined nucleon as well).

\begin{figure}[!!htb]
\centering
\begin{tabular}{l}
\includegraphics[width=0.55\textwidth]{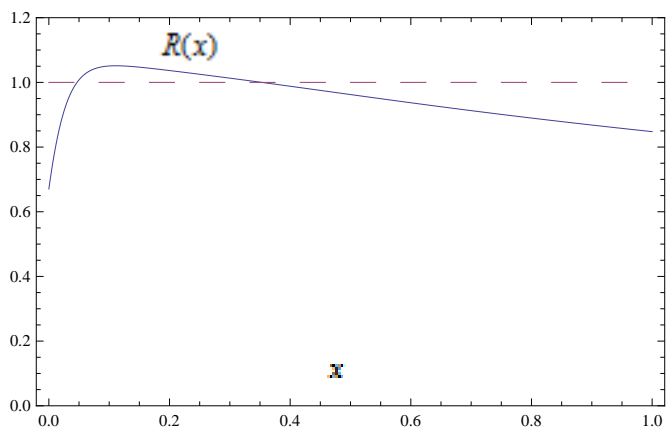}
\end{tabular}
\caption{The ratio of nuclear structure function R(x)(iron $^{55}$Fe to deuterium)}
  \label{fig:fig4}
\end{figure}

A nucleon in nuclei is affected inevitably by other nucleons in the nuclei and hence the equilibrium among valence quarks,
sea quarks and gluons in a free nucleon is broken down and a new equilibrium state is achieved, so the key parameters
$\xi_{v}$, $\xi_{s}$ and $\alpha$ will vary. Denoting the corresponding parameters for nucleon
in nuclei as $\xi_{v}^{'}$, $\xi_{s}^{'}$ and $\alpha^{'}$, we have taken $\xi_{v}=0.4$, $\xi_{s}=0.14$ and $\alpha=0.36$ for the free nucleon in the above
discussions, if taking $\xi_{v}^{'}=0.39$, $\xi_{s}^{'}=0.20$ and $\alpha^{'}=0.40$, we find R(x) for iron $^{55}$Fe fit the experimental data well
in the range of $x\in[0,0.7]$, as shown in Fig.~\ref{fig:fig4}.

%%%=============================================================
%%%%==============================================================

In the range of $x\in[0.7,1]$, experimental data show an apparent rise as $x$ increases, our curve depart from them and monotonously go down,
the reason is that we have neglected the effect of Fermi motion of the nucleons. Considering the influence of Fermi motion, Rozynek and Wilk
found that the trend of nucleonic structure function inside nucleus will change substantially in the high-$x$ direction\cite{Rozynek}.

Of course, nuclear environment is different for different nuclei, so the corresponding parameters $\xi_{v}^{'}$, $\xi_{s}^{'}$ and $\alpha^{'}$ will
vary with nuclei. This change is small, and it leads to the shape of ratio R(x) curve is a little bit different for different nuclei,
but the total variation tendency is alike.

\section{Summary} \label{sec:sum}
In this paper, we use a statistic model to calculate the parton distribution in a free nucleon and in a confined nucleon(nuclear EMC effect).
Though this model has not been able to explain the parton distribution over all $x$, it fit the experimental data well in most range of $x$ at
least reveal statistical law play an important role in the parton distribution of nucleon. Considering other effects like nucleon Fermi motion
and possible surface effects of detailed nuclear structure, one may give a better description of nucleon parton distribution over more $x$ and more
nuclei. All of the effort will be worth, for it help us to understand the internal physics of nucleon.

%%%--=================================================================
%%%=====            Acknowledgements        ==========================
%%5===================================================================

\begin{acknowledgments}
The author would to thank
Dr. Ming-Zhen Zhou and Dr. Wen-Long Sang for valuable discussions.
This work is partly supported by the National Natural Science Foundation of
 China under Grant No.11047028 and the Fundamental
 Research Funds of the Central Universities, Grant Number
 XDJK2012C040.
\end{acknowledgments}

%%====================================================================
%%%%%%%%%%%%%%%%%%%%    References  ==================================
%%%%==================================================================

\end{document}